# Implications of Charge Ordering in the High $T_c$ Cuprate Superconductors in the Far-infrared Spectroscopy

Y. H. Kim[1*] and P. H. Hor[2**]

[1]Department of Physics, University of Cincinnati, Cincinnati, Ohio 45221-0011, U. S. A.
[2]Department of Physics and Texas Center for Superconductivity
University of Houston, Houston, Texas 77204-5005, U. S. A.

(Dated: March 18, 2013)

We addressed the issue of the absence of the far-infrared signatures pertaining to charge ordering in the published far-infrared reflectivity data of $La_{2-x}Sr_xCuO_4$ single crystals while other experimental probes unravel that charge ordering is a hallmark of the superconducting cuprates. Through direct comparison of the far-infrared data reported by various groups side by side and also with the Raman scattering data, we found that the inconsistencies stem from the failure in capturing delicate spectral features embedded in the close-to-perfect *ab*-plane far-infrared reflectivity of $La_{2-x}Sr_xCuO_4$ single crystals by misidentifying the reflectivity as the Drude-like metallic reflectivity. The analysis of the close-to-true reflectivity data reveals that only a small fraction (< 3 %) of the total doping-induced charge carriers (electrons) are itinerant on the electron lattice made up with the rest of the electrons (> 97 %) at all doping levels up to 16 %. We conclude that the far-infrared reflectivity study is far from being ready to construct a coherent picture of the ubiquitous charge ordering phenomenon and its relationship with the high $T_c$ superconductivity.

## I. Introduction

The focus of the research of the high $T_c$ superconductivity in copper oxide superconductors (cuprates) is now zooming on identifying the physical origin of the energy gap in the normal state (pseudogap) of the doping-induced charge carriers (electrons) in the $CuO_2$ planes (*ab*-planes). Since the itinerant electrons do not have an energy gap unless the electrons are in a pair-bound state as in the BCS superconducting state, the presence of the energy gap of the electrons in the normal state has raised many intriguing questions and yet to date no consensus on the nature of the pseudogap state has been reached. This lack of consensus is particularly true of the experimental works mainly because of the inconsistencies in the measured physical quantities that should be independent of the experimental probes or laboratories. One prominent example was the case of the angle-resolved photoemission spectroscopy (ARPES) which is a powerful tool to obtain the momentum information of the electrons in the two-dimensional (2D) electronic structure. Through numerous ARPES work on superconducting cuprates, the momentum space picture of the electrons was constructed. However, because of the limitation in its energy resolution, there had been a disagreement on the doping dependency of the Fermi velocity of the electrons, for example, with that obtained through a low energy probe such as thermal conductivity measurement. Only after the ARPES technique was refined to use the 7 eV photons with an unprecedented energy resolution of ~ 3 meV, the doping dependence of the Fermi velocity extracted from the ARPES data[1] came to an qualitative agreement with the thermal conductivity data[2].

The same situation also arose between the ARPES and the scanning tunneling microscopy (STM) experiments. While STM studies consistently revealed prevailing evidences of the symmetry breaking electron ordered states in superconducting cuprates[3], only recently, through the continued improvement on lowering of the probe energy and on increasing the momentum and energy resolution, ARPES data began to unravel the true nature of the so-called "Fermi arc" which was known as the *d*-wave pairing energy gap of the electrons. In their photon-energy-dependent ARPES measurement, Hashimoto *et al.*[4] were able to resolve the *d*-wave-like nodal gap into two energy gaps after reaching the lowest energy with the highest resolution. Further ARPES studies finally drew a conclusion that the Fermi arc is not composed of the free electrons[5] but actually gapped[6]. This realization suggests that the Fermi arc has its origin in the broadened *localized* spectral weight along the nodal direction, implying that the electrons are in the *localized* state that has been the main result of the STM work.

Such overwhelming experimental evidence for the electron ordering was found not only through the surface sensitive probes such as STM and ARPES but also through the bulk measurements such as Nernst effect[7], resonant soft x-ray scattering[8], high energy x-ray diffraction[9], and magnetoresistance[10], which clearly demonstrate that the electron ordering is an intrinsic bulk property of the superconducting cuprates. However, it is peculiar to note that the far-infrared (far-IR) spectroscopy, which should be the most favorable probe to study the low-lying electronic excitations in superconducting cuprates, has not been able to consistently produce a definite signature of the symmetry-breaking electron ordered state. Motivated by this puzzle, we investigated the far-IR reflectivity data of $La_{2-x}Sr_xCuO_4$ (LSCO) system reported by various groups in the past two decades. Through direct comparison of the data side by side we found that no two *ab*-plane far-IR reflectivities published by different groups are in agreement with each other for frequencies ($\omega$) below 150 cm$^{-1}$. On the contrary, however, we notice that the *c*-axis (polarization perpendicular to the *ab*-planes) far-IR reflectivity of LSCO measured by many different groups are more or less in agreement with each other at all frequencies[11].

The main difference between the *c*-axis and *ab*-plane reflectivity measurements lies in the fact that the *c*-axis far-IR reflectivity of LSCO is flat at the level of ~ 40 % for $\omega < 200$ cm$^{-1}$ with very little temperature (*T*) dependence except for the emergence of the narrow Josephson plasma resonance dip in the reflectivity at $T \leq T_c$ while the *ab*-plane reflectivity exceeds 90 % for $\omega < 200$ cm$^{-1}$ at $T < 200$ K. Therefore, from the far-IR experimental point of view, the source of the problem may not be in the reflectivity measurement itself since all the groups reported a similar *c*-axis far-IR reflectivity despite the fact that the *c*-axis far-IR reflectivity should be noisier to measure especially for $\omega < 100$ cm$^{-1}$ and, moreover, applying a polarizer would further decrease the signal to noise ratio. But rather the problem might have arisen during the data analysis by losing the spectral information embedded in the close-to-perfect *ab*-plane reflectivity of LSCO because the high *ab*-plane reflectivity of LSCO that approaches 100 % in the far-IR, as we will show below, does not come from the electrons in a Drude-like metal.

## II. Review of the Far-IR Data

In order to elucidate the far-IR reflectivity problem, we compiled, as an example, the *T*-dependence of the real part of the *ab*-plane far-IR conductivities ($\sigma_1$) of LSCO single crystals at two different doping concentrations (x ~ 0.12 and x ~ 0.16) published by five different groups and present them in Figure 1. The left column shows the *T*-dependence of the underdoped La$_2$CuO$_{4.06}$ ($T_c$ = 36 K)[12] and LSCO near x ~ 0.12[13-16]. There are dramatic differences that are readily noticeable for $\omega < 100$ cm$^{-1}$. For instance, in Figure 1(a) there develops a peak at around ~ 100 cm$^{-1}$ at room *T*. As *T* decreases, its position shifts towards the lower frequencies and its strength substantially increases even in the superconducting state[12], indicating that the far-IR tail in $\sigma_1$ is not due to the free carrier conductivity. In Figure 1(b), there emerges a sharp intense peak at ~ 30 cm$^{-1}$ out of a broad background conductivity as *T* decreases[13]. Figure 1(c) shows that there are two spectral contributions: one is a weakly $T$-dependent broad peak centered at ~ 100 cm$^{-1}$ and the other is like the tail of a Drude-like $\sigma_1$ which increases its strength in the normal state ($T > T_c$) with decreasing $T$ but partially loses its strength in the superconducting state ($T < T_c$)[14]. In contrast, the free carrier spectral weight in Figure 1(d) seems to fully disappear in the superconducting state, leaving a broad background conductivity peak centered at around ~ 90 cm$^{-1}$ behind[15]. In Figure 1(e), a broad peak at ~ 150 cm$^{-1}$ at 300 K grows its strength with decreasing $T$ while its position becomes red-shifted to ~ 110 cm$^{-1}$ as $T$ decreases[16]. Furthermore this peak is dominating the entire spectra at all $T$ and the itinerant electron contribution is extremely small when compared with the intense peak[16].

In the right column of Figure 1, the *ab*-plane far-IR $\sigma_1$ of the nearly optimally or optimally doped LSCO (x = 0.16) is shown. Figure 1(f) shows rather flat conductivity for both normal and superconducting state with a small bump at ~ 100 cm$^{-1}$ for x = 0.14 LSCO[17]. In Figure 1(g), there are at least four peaks develop including one at ~ 100 cm$^{-1}$ as $T$ decreases even below $T_c$[13]. Figure 1(h) shows two contributions (the broad peak at ~ 100 cm$^{-1}$ and the free carrier behavior seen at x = 0.12)[14]. However, its Drude-like tail grows sharper and even stronger at $T$ = 5 K ($T_c$ ~ 38 K)[14]. Figure 1(i) shows that the conductivity is dominated by the peak centered at ~ 120 cm$^{-1}$ which reaches its maximum strength at $T < T_c$ while its position remains unchanged with decreasing $T$[18]. A similar trend as Figure 1(e) for x = 0.11 LSCO is seen in Figure 1(j) for x = 0.16 LSCO except that the position is now shifted to ~ 150 cm$^{-1}$ and again this peak is dominating the entire conductivity at all $T$[19]. In this Figure, qualitatively we can see that there exists a substantial spectral weight in the $w$ range between ~50 cm$^{-1}$ and ~150 cm$^{-1}$ even $T < T_c$.

In order to shed more light, the doping dependence of the far-IR *ab*-plane $\sigma_1$ obtained by two groups was compared to sort out the progressive spectral changes upon doping. The displayed data in Figure 2 are chosen for $T$ well below $T_c$ to insure that the free electron contribution to the ab-plane $\sigma_1$ has been transferred to a delta function centered at $\omega = 0$ in the superconducting state. The doping-dependent ab-plane far-IR $\sigma_1$ of LSCO reported by Padilla et al.[20] and by Dumm et al.[15] are displayed in the left column of Figure 2 and those obtained by Kim et al.[16, 19] in the right column. These two studies are chosen because they exhibit an internal consistency upon doping. It is clear that there develops a broad structure upon doping which is centered at around ~ 90 cm$^{-1}$ in the data shown in the left column whereas the broad structure appears at around ~ 110 cm$^{-1}$ at the lower doping and at ~ 150 cm$^{-1}$ at optimal doping in the right column. Naturally arising question is then which far-IR study correctly reflects the far-IR charge dynamics of LSCO?

The answer may be found in the Raman scattering data. In the Raman scattering studies of LSCO[21,22] at various doping concentrations, it was found that there appear two new ab-plane Raman modes, one at ~ 115 cm$^{-1}$ ($S_1$) and the other at ~ 150 cm$^{-1}$ ($S_2$) at 3 % Sr-doping as shown in the bottom panel of the left column of Figure 3. Upon increasing doping to 12.5 % and to 17 %, the $S_2$-mode continues to grow its intensity while the $S_1$-mode continues to decrease as doping increases (see the left column of Figure 3). In fact it was shown that the intensity of the $S_2$-mode reaches its maximum at 16 % Sr-doping[23]. Finally at 24 % Sr-doping in which the HTS is about to cease to exist, the $S_1$-mode now disappears and also the $S_2$-mode substantially loses its strength as shown in the top panel of the left column. Hence, when this Raman data is compared with the ab-plane far-IR $\sigma_1$ of LSCO single crystals obtained by Kim et al.[16,19] (see the right column of Figure 3), it is clear that the broad intense peak observed in the far-IR at ~ 110 cm$^{-1}$ (~ 150 cm$^{-1}$ at 16 % Sr-doping) is a superposition of the $S_1$ and $S_2$ modes seen in the Raman scattering as their frequencies and linewidth suggest. Therefore, since the $S_1$ and $S_2$ modes are independent of the oxygen isotope substitute[23], we conclude that the far-IR mode at $\omega \sim 110$ cm$^{-1}$ – 150 cm$^{-1}$ has its origin in the phonons which are highly *ab*-plane polarized bending motion of the La/Sr atoms attached to the apical oxygen of the octahedron cage[22,24]. The charge symmetry of this Raman-active mode is broken upon doping due to the strong electron-lattice coupling of the electrons to the CuO$_2$ lattice to make the Raman-active mode infrared active[25]. Thus the intensity of the infrared-active vibration (IRAV) mode is proportional to the density of the electrons that are coupled to the lattice.

The appearance of the IRAV modes in the Raman scattering as soft-modes (S$_1$ and S$_2$ modes) suggests that the inversion symmetry is broken by the formation of the electron ordered structures[21,23]. This has a profound consequence in the electron Raman data analysis because not only the intensity of the S$_2$ mode grows with decreasing the temperature but also its width rapidly broadens below 100 K[21] as *T* decreases. For example, the apparent new structure at the frequency of S$_2$ in the electron Raman data is due to the incomplete compensation when the phonons are subtracted from the Raman data to get the electronic background information as a function of $T$[26]. Hence the so-called B$_{1g}$ and B$_{2g}$ modes in the electron Raman data of LSCO have their origin in the soft S$_1$ and S$_2$ modes. Nevertheless the interpretation thereof correctly assessed that the electron lattice (EL) formation is responsible for the B$_{1g}$ (S$_1$) and B$_{2g}$ (S$_2$) modes[27,28].

### III. Far-IR Data Analysis and Discussion

Since we have shown that our data is intrinsically consistent with the Raman data in the far IR regime, in Figure 4 we present the close-to-true *ab*-plane reflectivity data of LSCO single crystals of x = 0.063, 0.07, 0.09, 0.11, and 0.16 measured at *T* = 8 K (*T* = 14 K for x = 0.16 sample). All the reflectivity data are plotted 'as taken' except for $\omega \leq 13$ cm$^{-1}$ for all the underdoped samples and for $\omega < 25$ cm$^{-1}$ for x = 0.16 sample where the smoothing of the data was inevitable. The angle of incidence was 8º. There are two things readily noticeable: (1) The far-IR reflectivity of LSCO never reaches 100 % even in the superconducting state except for the frequencies below ~ 15 cm$^{-1}$ within the experimental uncertainty and (2) there is a weak broad bump at ~ 150 cm$^{-1}$ that mildly decreases toward a local minimum as $w$ decreases. This makes the *ab*-plane far-IR reflectivity measurement of LSCO single crystals nontrivial because such a small feature in the reflectivity manifests as an intense peak in the far-IR $\sigma_1$ because of its presence on top of the over 90 % background reflectivity. This high far-IR *ab*-plane reflectivity of LSCO arises from the strong and rapidly varying absorption features which involve a large imaginary part of the index of refraction[16]. Thus the angle of incidence of the far-IR beam on the sample surface is critically important not because of the possible leakage of the *c*-axis spectral information into the *ab*-plane far-IR reflectivity as commonly argued but because of the highly *ab*-plane polarized absorption features that are present for $\omega < 200$ cm$^{-1}$. Since the *c*-axis far-IR reflectivity is featureless for $\omega < 200$ cm$^{-1}$ at the level of ~ 40 – 50 %, it is safe to conclude that all the absorption features observed in the far-IR reflectivity of LSCO for $\omega < 200$ cm$^{-1}$ originate from the *ab*-plane charge dynamics. The same is also true in the polycrystalline sample spectra[19].

It is thus now clear that the inconsistency in the published *ab*-plane far-IR reflectivity depends wholly on whether or not these delicate features are captured. We found that when the angle of incidence is large as ~ 15º (typically ~ 12º - 15º for a commercial reflectivity measurement set-up), the broad weak spectral features in the reflectivity can easily be wiped out[16]. Furthermore, since the far-IR spectroscopy utilizes Mylar beam splitters of various thicknesses in conjunction with cold/warm filters in order to maximize the detector sensitivity at the given frequency range, it is necessary to merge a series of reflectivity data sets for the Kramers-Kronig analysis. Therefore, the linearity of the detector with high sensitivity must be insured and data smoothing must be avoided. The data presented in this work and all other far-IR data published by the present authors were not smoothened unless specified. The frequency overlapping regions (typically 10 cm$^{-1}$ < $\Delta\omega$ < 30 cm$^{-1}$) between two different data sets were averaged in the process of merging. As can be seen in Figure 4, a few delicate weak structures appear on top of the background reflectivity that reaches above 90 %. These structures can easily be missed by a first-smoothing-then-merging procedure. In addition the local minimum in the reflectivity ($\omega$ < 50 cm$^{-1}$), which leads to an upturn toward 100 % reflectivity as $\omega \rightarrow 0$, can easily be disregarded as experimental uncertainty when the signal-to-noise ratio is not good enough.

Another serious consequence of failing in capturing the delicate structures in the far-IR reflectivity is leaving out the EL collective modes. While Kim *et al.* observed a series of EL collective modes at ~ 23 cm$^{-1}$ (~ 3 meV), ~ 46 cm$^{-1}$ (~ 6 meV), and ~ 72 cm$^{-1}$ (~ 9 meV) both in the underdoped and optimally doped polycrystalline[29,30], which is free from the reflectivity measurement problem, and single crystalline[16,19] LSCO samples, there has been no other far-IR work that reported an observation of the EL collective modes in LSCO (incidentally these EL collective modes do not have their counter part Raman-active modes as in the case of the IRAV modes). We suggest that Raman scattering measurement should also be able to detect the EL collective modes through the inversion symmetry breaking although the instrumental limitation may be a major factor for LSCO system because of their low frequencies. However, we believe that one of the EL collective modes of YBa$_2$Cu$_3$O$_y$ was observed in the electron Raman scattering study which revealed the presence of the mode at ~ 240 cm$^{-1}$ (~ 30 meV) in y = 6.54 (T$_c$ = 54 K) sample[28]. We point out that this mode corresponds to the EL collective mode observed in the far-IR at ~ 220 cm$^{-1}$ in YBa$_2$Cu$_3$O$_{6.5}$[31] which is also close in energy to the neutron resonance of the same sample measured at ~ 30 meV for $T \leq T_c = 60$ K[32]. Therefore, the presence of the EL collective modes and their relationship with the neutron resonances[33] remain to be confirmed through careful far-IR and Raman scattering experiments. In particular, observation of the EL collective modes at ~ 330 cm$^{-1}$ (41 meV) in YBa$_2$Cu$_3$O$_7$ ($T_c$ = 90 K) and at ~ 110 cm$^{-1}$ (~ 14 meV) in YBa$_2$Cu$_3$O$_{6+y}$ ($T_c$ = 30 K) through the far-IR and Raman scattering measurements would definitely help to establish the relationship between the charge channel and the spin channel in high $T_c$ superconductivity[33].

The Kramers-Kronig derived *ab*-plane $\varepsilon_1(\omega)$ of LSCO in the superconducting state is shown in Figure 5 at various doping levels. It is important to note that the free electron contribution, which is characterized by the negative $\varepsilon_1(\omega)$ for $\omega$ below the screened plasma frequency ($\tilde{\omega}_p$), is confined to $\omega < 100$ cm$^{-1}$ at all doping levels up to 16%. Further insight into the dynamics of the itinerant electrons may be obtained by adopting the two-fluid model $\varepsilon(\omega) = \varepsilon_\infty - f_s \omega_p^2/\omega^2 - (1-f_s)\omega_p^2/\omega(\omega + i\Gamma)$ for the itinerant electrons in the superconducting state where $f_s$ is the superfluid fraction which is typically expressed as $\omega_{ps}^2/\omega_{pn}^2$ in other works; $\omega_p$ is the unscreened plasma frequency given by $\omega_p^2 = ne^2/\pi m_e c^2$ ($n$ = electron density and $m_e$ = electron mass); $\varepsilon_\infty$ is the dielectric constant of the medium (*i.e.*, $\varepsilon_\infty$ = 1 in empty space); and $\Gamma$ is the scattering rate in the normal state. Since $f_s \to 1$ when $T << T_c$, $\varepsilon(\omega)$ may be approximated as $\varepsilon(\omega) \approx \varepsilon_1(\omega) = \varepsilon_\infty - \omega_p^2/\omega^2$ for the data presented in this work. Thus, by plotting $\varepsilon_1(\omega)$ of the lowest $T$ data versus $\omega^{-2}$, one may obtain the information about $\varepsilon_\infty$ from the y-intercept and $\omega_p^2$ from the slope as displayed in Figure 6. As a crosscheck, $\omega_p^2$ may also be calculated from the imaginary part of the conductivity, $\sigma_2$ by taking the limit $\lim_{\omega \to 0} 60 \omega \sigma_2(\omega) = \omega_p^2$ and the results are displayed in Figure 7 along with the result found from the linear fit shown in Figure 6 (cyan square).

From the $\omega_p$ of the itinerant electrons found in this analysis, the ratio of the itinerant electron density to the total electrons density ($n_0$) may be estimated from $\omega_p^2/\omega_{p0}^2$ where $\omega_{p0}^2 = n_0 e^2/\pi m_e c^2$ since $n_0$ is given by the Sr concentration and $m_e$ should be independent of the electron density. As summarized in Table 1, the fraction of the free electrons is only on the order of 1 % of the total electrons in underdoped LSCO and it reaches only 2.8 % at the optimal doping. Moreover the $\varepsilon_\infty$ that is responsible for the screening of the itinerant electrons is substantially large enough to push the $\tilde{\omega}_p = \omega_p/\sqrt{\varepsilon_\infty}$ down below 100 cm$^{-1}$ at all doping levels up to x = 0.16 (see Figure 5). This observation is significant because it implies that the itinerant electrons (< 3 %) in cupartes are present in a medium of gigantic $\varepsilon_\infty$. Since the theoretical estimation[34] of $\varepsilon_\infty$ of the cuprates is typically in the range of 20 – 40, such a gigantic dielectric constant can be produced in cuprates only when the electrons are organized to form an EL whose $\varepsilon_1(\omega)$ takes the form given by $\varepsilon_1 = 1 - \Omega_p^2/(\omega^2 - \omega_G^2) \xrightarrow{\omega \to 0} \varepsilon_1(0) = 1 + \Omega_p^2/\omega_G^2$ where $\Omega_p^2 = n_{EL} e^2/\pi m^* c^2$ ($n_{EL}$ = the density of the localized electrons in the EL and $m^*$ = effective mass of the localized electron in EL) and $\omega_G$ is the EL collective mode frequency. Thus such massive screening of the itinerant electrons is a clear indication that the itinerant electrons in LSCO are intimately coupled to the EL in such a way that the $\varepsilon_1(0)$ of the EL serves as $\varepsilon_\infty$ for the itinerant electrons. This is in stark contrast to the case of conventional charge-density-wave (CDW) systems where the CDW order competes with the metallic phase[35-37]. It was observed that the $\tilde{\omega}_p$ of the electrons in the metallic phase of 1D CDW[35] or 2D CDW system[36,37] remains unchanged above and below the CDW transition, which is a clear indication that the electrons in the metallic phase of these systems are decoupled from the CDW in the normal state. This observation led us to conclude that the EL order in cuprate superconductors cannot be competing with the itinerant electrons since otherwise such massive screening of the itinerant electrons would not have been possible. Therefore the itinerant electrons in LSCO are supported and screened by the EL formed by the rest of the electrons.

## IV. Summary and Conclusion

In this paper we addressed the inconsistencies in the far-IR reflectivity in the published *ab*-plane reflectivity data of LSCO single crystals based on the observation that no two *ab*-plane far-IR reflectivities reported by different groups are in agreement with each other for $\omega < 150$ cm$^{-1}$ in LSCO. Consequently this problem caused the total lack of consensus on the far-IR dynamics of the electrons in cuprates and the failure to observe the electron ordering signatures in the far-IR *ab*-plane reflectivity. We identified that the source of the problem was in the misidentification of the close-to-perfect *ab*-plane far-IR reflectivity of LSCO as the reflectivity of a Drude-like metal. Through direct comparison with the Raman scattering work on LSCO the close-to-true far-IR reflectivity data was determined and subsequent analysis of the reflectivity data, we found that only a small fraction (< 3 %) of $n_0$ are itinerant on the electron lattice formed by the *localized* electrons (> 97 % of $n_0$) as evidenced by the massive screening of the itinerant electrons. Therefore far-IR charge dynamics study of the high $T_c$ cuprates is not complete contrary to the popular view and more careful far-IR and Raman scattering measurements are required in order to bring about a coherent picture of the ubiquitous electron lattice phase and its topology seen in STM and other experiments.

**Table 1.** Unscreened plasma frequency ($\omega_p$) and the background dielectric constant ($\varepsilon_\infty$) extracted from the *ab*-plane $\varepsilon_1$ versus $\omega^{-2}$ plot of LSCO as well as the itinerant electron fraction (see the text) at various doping levels.

| $La_{2-x}Sr_xCuO_4$ | $\omega_p (cm^{-1})$ | $\varepsilon_\infty$ | $n/n_0$ |
|---|---|---|---|
| x = 0.063 | ~ 540 ± 20 | ~ 560 ± 20 | ~ 0.005 |
| x = 0.07 | ~ 780 ± 20 | ~ 680 ± 20 | ~ 0.009 |
| x = 0.09 | ~ 860 ± 40 | ~ 2100 ± 60 | ~ 0.01 |
| x = 0.11 | ~ 1400 ± 40 | ~ 1400 ± 40 | ~ 0.02 |
| x = 0.16 | ~ 2000 ± 40 | ~ 1600 ± 40 | ~ 0.028 |

**Figure Captions**

Figure 1. Temperature dependence of *ab*-plane $\sigma_1$ of LSCO single crystals reported by various groups. *Left column*: (a) $T = 4$ K (blue), 25 K (cyan), 100 K (magenta), and 150 K (orange) [Ref. 12]; (b) $T = 30$ K (green), 100 K (magenta), and 300 K (red) [Ref. 13]; (c) $T = 5$ K (blue), 50 K (green), 200 K (orange), and 300 K (red) [Ref. 14]; (d) $T = 5$ K (blue), 24 K (green), and 32 K (green) [Ref. 15]; and (e) $T = 8$ K (blue), 30 K (green), 100 K (magenta), 200 K (orange), and 300 K (red) [Ref. 16]. *Right column*: (f) $T = 10$ K (blue) and 36 K (green) [Ref. 17]; (g) $T = 30$ K (green), 100 K (magenta), and 300 K (red) [Ref. 13]; (h) $T = 5$ K (blue), 20 K (cyan), 50 K (green), 200 K (orange), and 300 K (red) [Ref. 14]; (i) $T = 10$ K (blue), 60 K (green), and 150 K (orange) [Ref. 18]; and (j) $T = 14$ K (blue), 20 K (cyan), 50 K (green), 100 K (megenta), 200 K (orange), and 300 K (red) [Ref. 19].

Figure 2. Comparison of the doping dependence of the *ab*-plane $\sigma_1$ of LSCO single crystals reported by two different groups. The top panel of the left column is from Ref. 15 and the rest of the left column from Ref. 20. The top panel of the right column is taken from Ref. 19 and the rest from Ref. 16.

Figure 3. Direct comparison of the $\sigma_1$ data of LSCO at $T = 8$ K ($T = 14$ K for x = 0.16) of Kim *et al.* (Refs. 16 and 19) with the Raman scattering data ($T = 10$ K) measured by Lampakis *et al.* (Ref. 22).

Figure 4. Far-infrared reflectivity of the *ab*-plane of LSCO single crystals at various doping levels at $T = 8$ K (Ref. 16) and at $T = 14$ K for x = 0.16 (Ref. 16 and 19).

Figure 5. Real part of the dielectric function ($\varepsilon_1$) of the electrons in the *ab*-plane of LSCO calculated from the reflectivity shown in Figure 4. Notice that the negative region below the screened plasma frequency ($\tilde{\omega}_p$), which belong to the itinerant electrons is confined to the region $\omega < 100$ cm$^{-1}$.

Figure 6. $\varepsilon_1$ versus $\omega^{-2}$ plot at different doping levels. The slope gives $\omega_p^2$ and *y*-intercept gives

the value of $\varepsilon_\infty$. The results are summarized in Table 1.

Figure 7. The graph of $60\omega\sigma_2(\omega)$ versus $\omega$ at five doping levels. The $\omega_p^2$ obtained from the slope of the graph shown in Figure 6 is also indicated (cyan square).

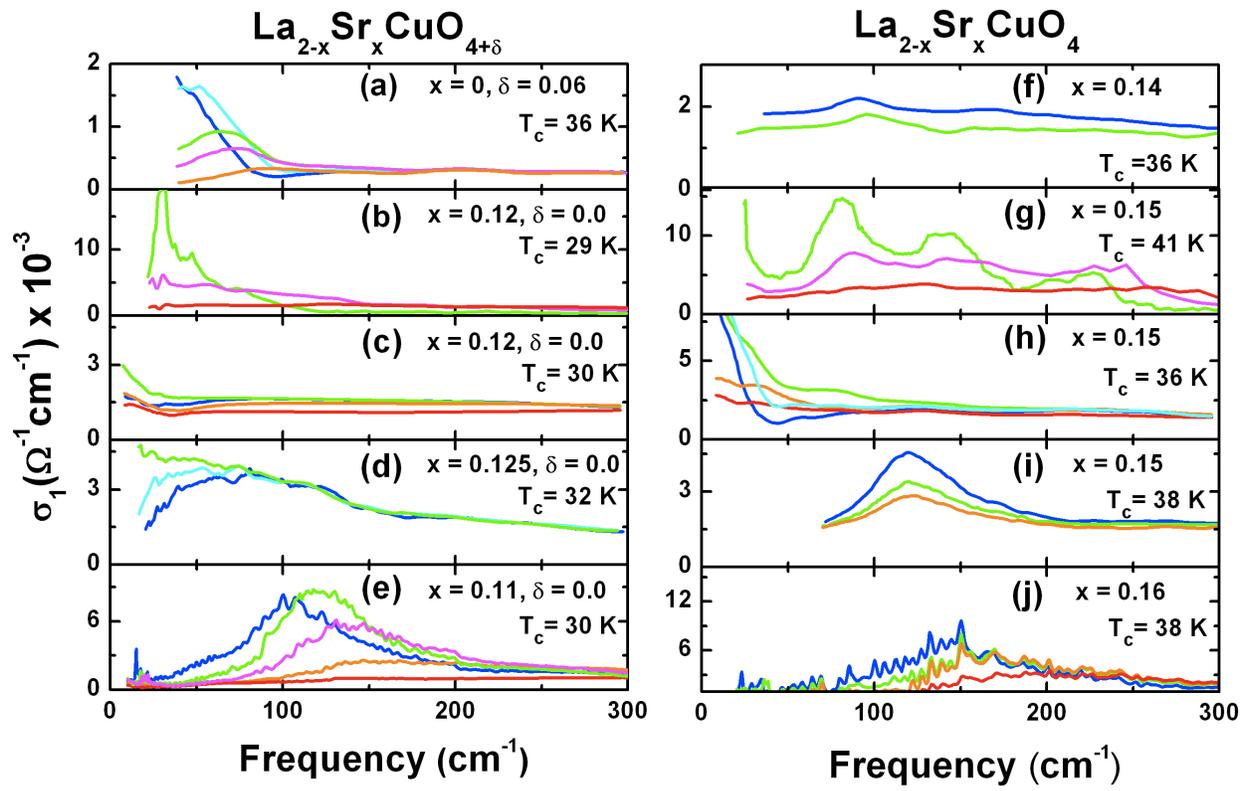

Figure 1

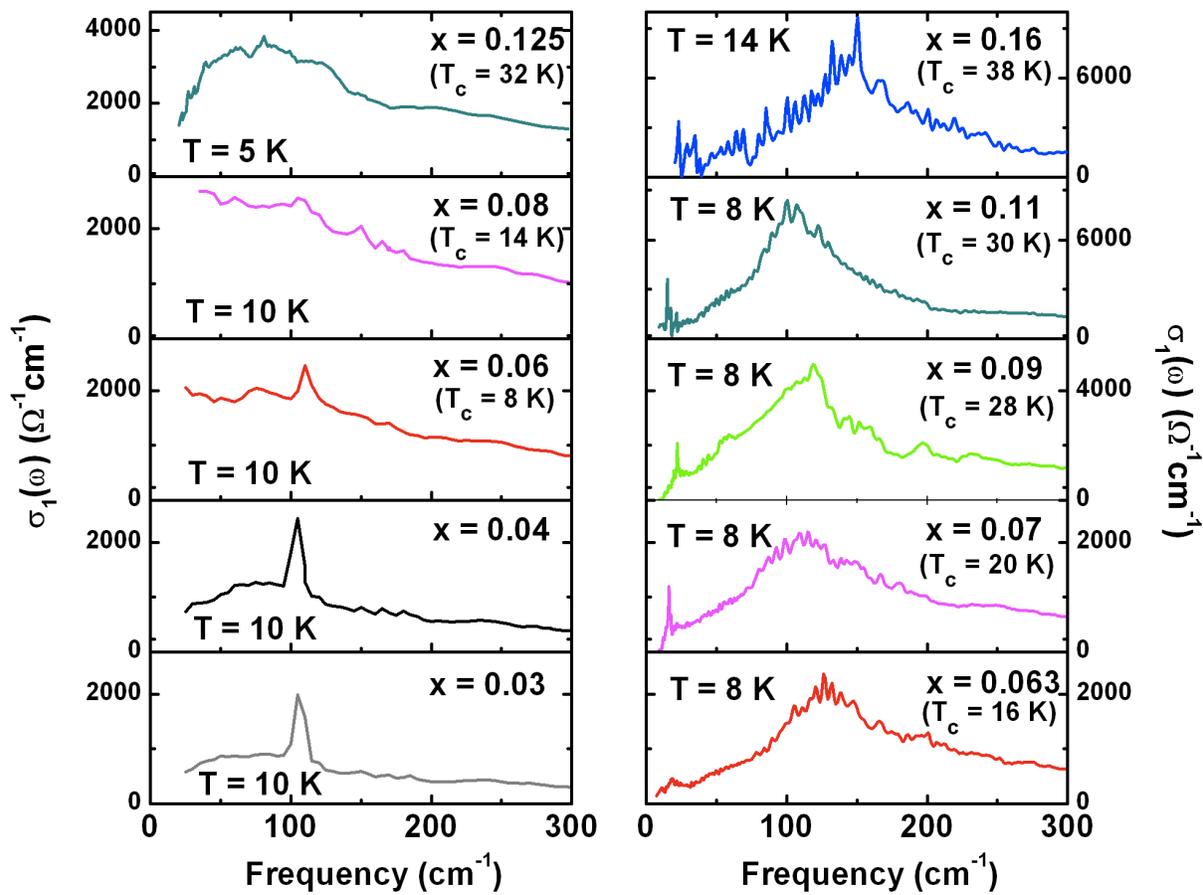

Figure 2

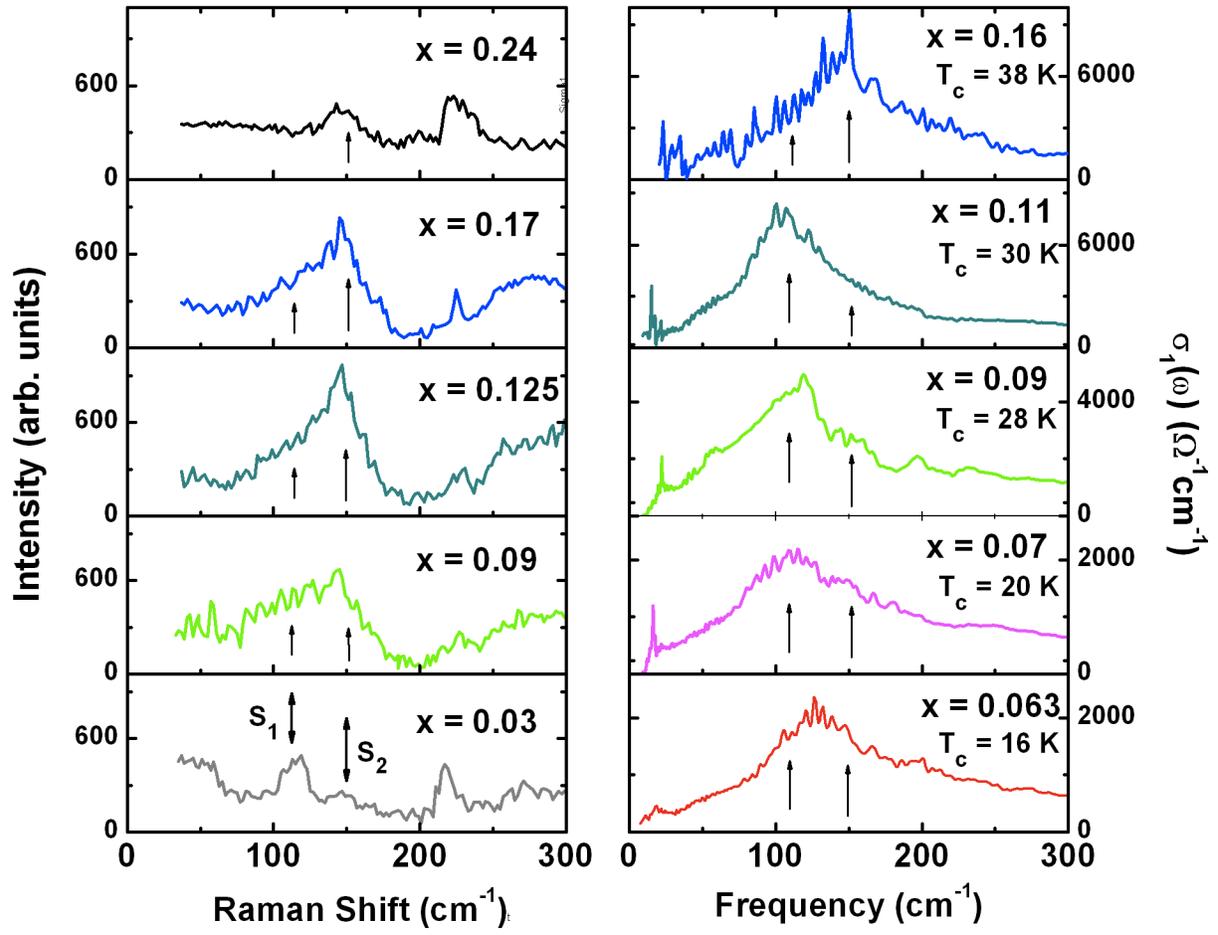

**Figure 3**

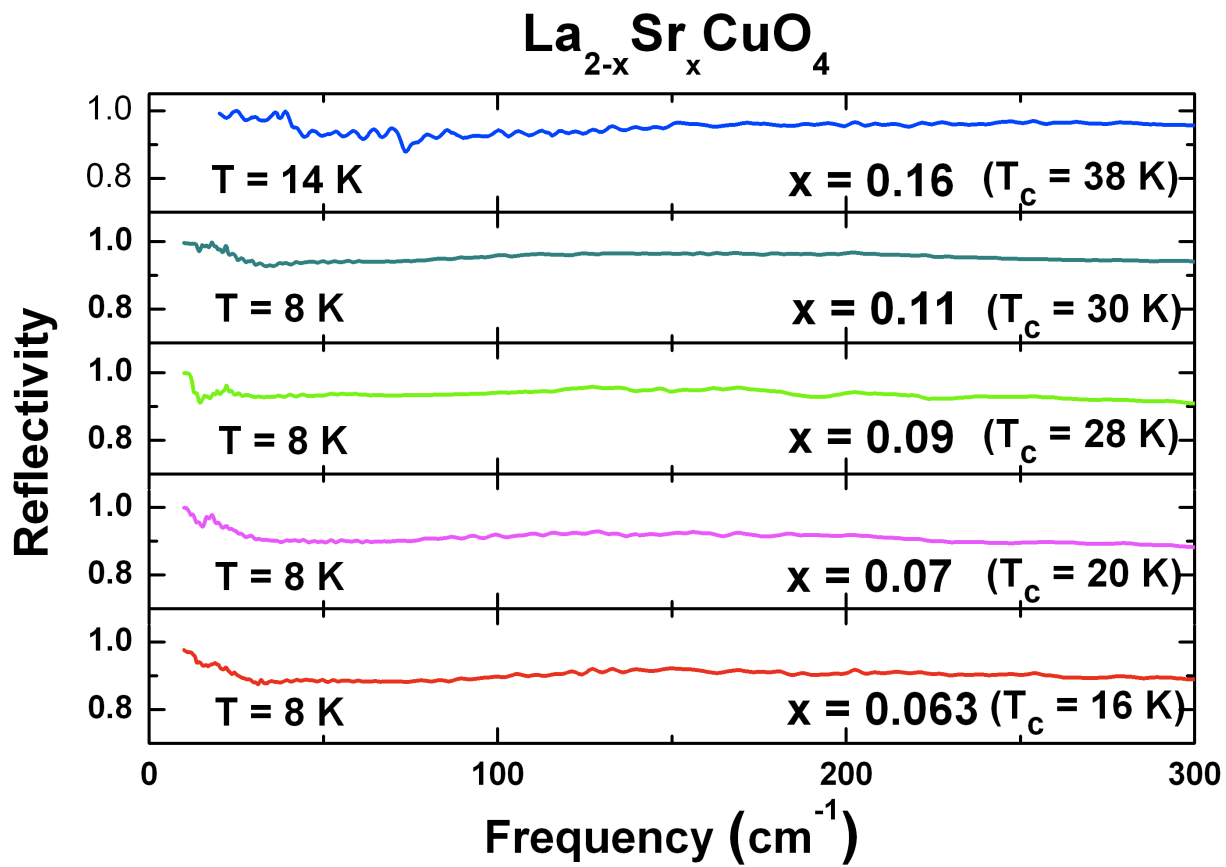

**Figure 4**

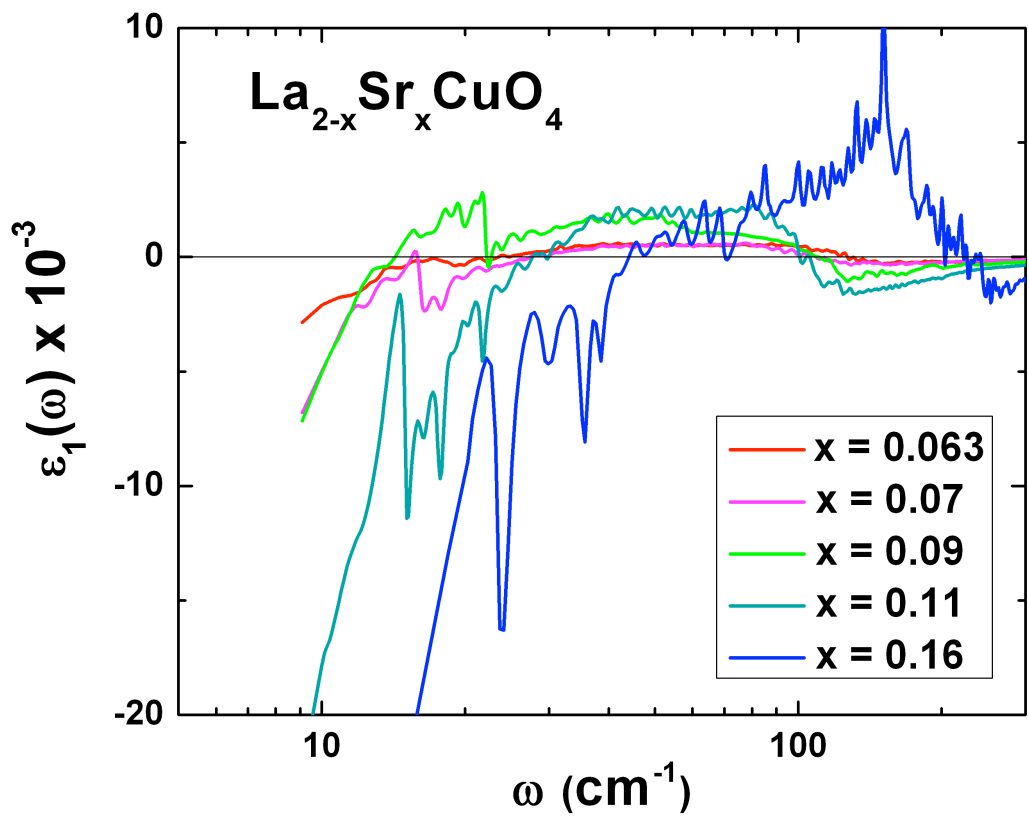

**Figure 5**

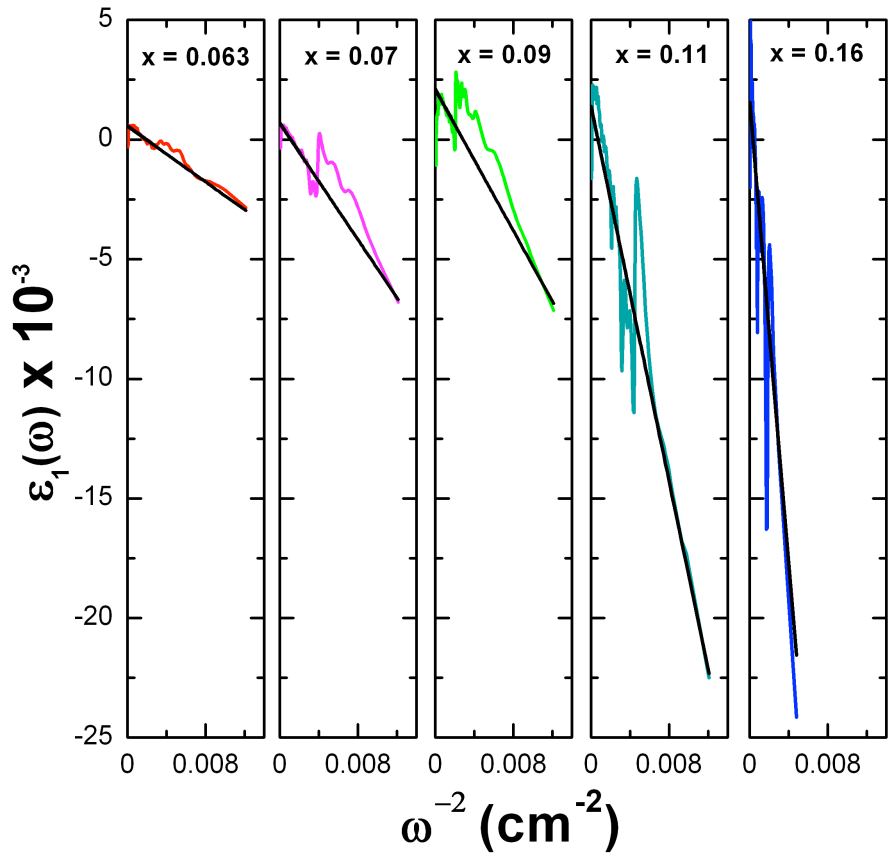

**Figure 6**

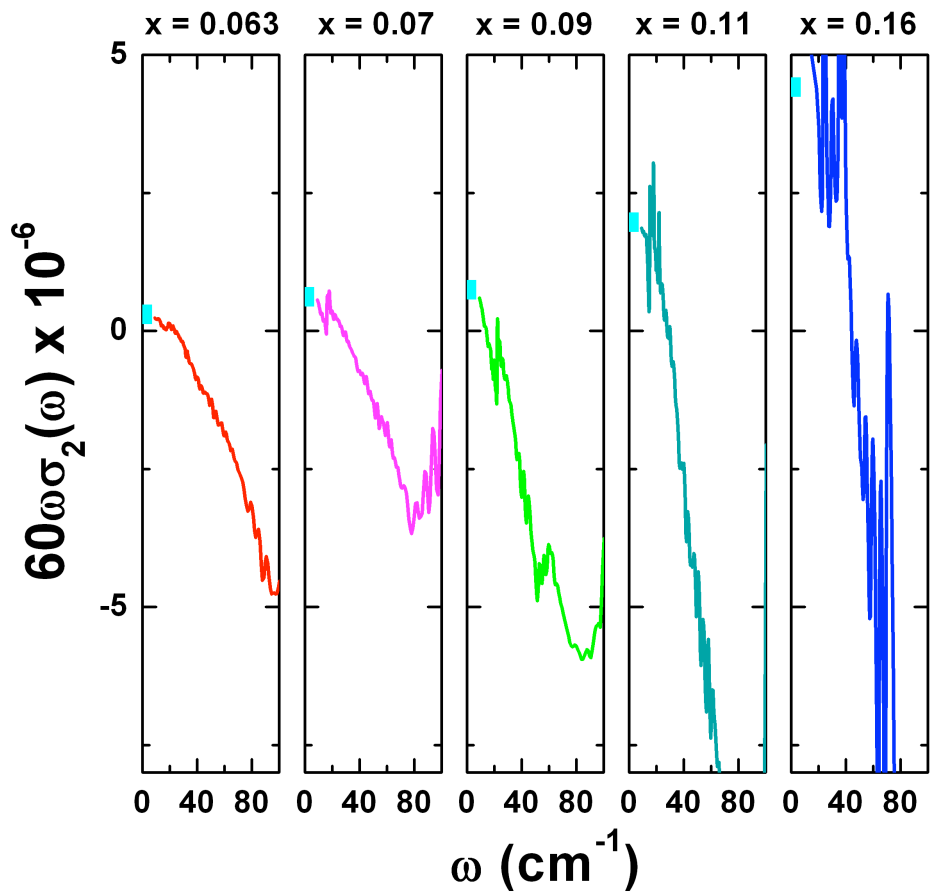

Figure 7